\documentclass[9pt, conference]{IEEEtran}
\IEEEoverridecommandlockouts
\usepackage{cite}
\usepackage{amsmath,amssymb,amsfonts}
\usepackage{algorithmic}
\usepackage{graphicx}
\usepackage{textcomp}
\usepackage{xcolor}
\usepackage{booktabs}
\usepackage{multirow}
\usepackage{diagbox}

\usepackage[marginal]{footmisc}

\def\BibTeX{{\rm B\kern-.05em{\sc i\kern-.025em b}\kern-.08em
    T\kern-.1667em\lower.7ex\hbox{E}\kern-.125emX}}
\begin{document}

\title{Metadata-Enhanced Speech Emotion Recognition: Augmented Residual Integration and Co-Attention in Two-Stage Fine-Tuning  \\
\thanks{$^\#$Equal contribution,  
 \IEEEauthorrefmark{1}Corresponding author}
\thanks{This work was supported by the National Natural Science Foundation of China under Grant No. 62276153, and by a grant from the Tsinghua University-Tsingshang Joint Institute for Smart Scene Innovation Design.}
}

\author{\IEEEauthorblockN{Zixiang Wan$^1$$^\#$, Ziyue Qiu$^2$$^\#$, Yiyang Liu$^3$, Wei-Qiang Zhang$^4$\IEEEauthorrefmark{1}}
\IEEEauthorblockA{
$^1$School of Computer Science, Beijing University of Posts and Telecommunications, Beijing, China \\
$^2$Ira A. Fulton Schools of Engineering, Arizona State University, Tempe, United States \\
$^3$Xingjian College, Tsinghua University, Beijing 100084, China \\
$^4$Department of Electronic Engineering, Tsinghua University, Beijing 100084, China \\
\tt wzx802@bupt.edu.cn, zqiu30@asu.edu, wqzhang@tsinghua.edu.cn}}

\maketitle
\begin{abstract}
Speech Emotion Recognition (SER) involves analyzing vocal expressions to determine the emotional state of speakers, where the comprehensive and thorough utilization of audio information is paramount. Therefore, we propose a novel approach on self-supervised learning (SSL) models that employs all available auxiliary information—specifically metadata—to enhance performance. Through a two-stage fine-tuning method in multi-task learning, we introduce the Augmented Residual Integration (ARI) module, which enhances transformer layers in encoder of SSL models. The module efficiently preserves acoustic features across all different levels, thereby significantly improving the performance of metadata-related auxiliary tasks that require various levels of features. Moreover, the Co-attention module is incorporated due to its complementary nature with ARI, enabling the model to effectively utilize multidimensional information and contextual relationships from metadata-related auxiliary tasks. Under pre-trained base models and speaker-independent setup, our approach consistently surpasses state-of-the-art (SOTA) models on multiple SSL encoders for the IEMOCAP dataset.

\end{abstract}

\begin{IEEEkeywords}
speech emotion recognition, metadata, multi-task learning, self-supervised learning model, fine-tuning
\end{IEEEkeywords}

\section{Introduction}
Speech Emotion Recognition (SER) is an advanced technology that uses artificial intelligence techniques to recognize emotional states in speech.
It has applications in areas \cite{zhang2021multilingual, morais2022speech, padi2022multimodal, twostageMTL,luo2022critical}  such as virtual assistants\cite{lu23d_interspeech}, customer service\cite{agrawal23b_interspeech}, and healthcare\cite{solinsky23_interspeech,10448004,10095522}. Studies have highlighted its importance in enhancing user experience and personalized services. 

However, since audio signal is a complex, non-linear, high-dimensional data type\cite{shaheen23_interspeech}, and emotional expression is conveyed through the combination and variation of multiple features instead of a single acoustic feature, it becomes challenging for AI models to identify clear patterns, which makes SER tasks particularly difficult.

Researchers have identified that incorporating auxiliary information, or metadata, into SER can significantly enhance its performance\cite{twostageMTL}, which reveals that effective SER requires a comprehensive understanding of the diverse factors that influence emotional expression in speech, including gender, speaker information, the style of speech, semantic content of speech etc\cite{li2022fusing,savchenko2021speaker,twostageMTL,schuller2004speech,ghosh2022mmer}. 

Thus, given that SER can benefit from this metadata, multi-task learning becomes natural choice\cite{hsu23_interspeech}. However, there are still challenges remaining in traditional MTL approaches\cite{10.1145/3372806.3372818, 8652301}. One challenge is the quality of auxiliary tasks if the auxiliary tasks are not accurate, they can lead to negative transfer, degrading the performance of the primary task\cite{ghosh23b_interspeech,Schuller2020}. 
Another challenge is that incorporating too many auxiliary tasks may overwhelm the model, leading to poor performance due to an inability to efficiently integrate high-dimensional information\cite{yu2020gradient,Valada2018}. 
Furthermore, the common practice of merging auxiliary tasks in a simplistic manner like weighted sum\cite{10447183} can lead to suboptimal use of the data, reducing overall model efficiency\cite{Kung2021}. 

To address the challenges in multi-task learning for Speech Emotion Recognition (SER), we introduce improvements within a two-stage fine-tuning approach on transformer-based self-supervised learning (SSL) models, which extract high-level speech features from raw audio and are fine-tuned with a small amount of labeled data. The main contributions of our work are summarized as follows:

\begin{itemize}

\item An Augmented Residual Integration (ARI) module is introduced to enhance the transformer-based encoder in SSL models, which effectively preserving multi-level acoustic feature suited for different metadata-related auxiliary tasks and outperforming the traditional weighted sum approach for every auxiliary tasks.
\item Co-attention module is novelly incorporated into the framework of two-stage fine-tuning. The module is superior at utilizing the multidimensional information from the ARI-enhanced encoder and the contextual relationships between all auxiliary tasks and their associated metadata.

\end{itemize}

The complementary nature of the two modules lies in their distinct strengths: the ARI module excels at contributing high-quality multidimensional information to auxiliary tasks by preserving features at various levels, while the Co-attention module efficiently utilizes this multidimensional information. This synergy enables the proposed approach to accommodate each type of available metadata (e.g., gender, speech style) as an auxiliary task or modality\cite{li2019towards}. The enhanced ability to process multimodal information extracted from metadata, with performance improvements positively correlating with the number of auxiliary tasks, is observed and validated on different transformer-based SSL models. Ultimately, under pre-trained base SSL models and a speaker-independent setup, our approach consistently outperforms state-of-the-art (SOTA) models on the IEMOCAP dataset.

\begin{figure*}[ht]  
  \centering  
  \includegraphics[width=\textwidth]{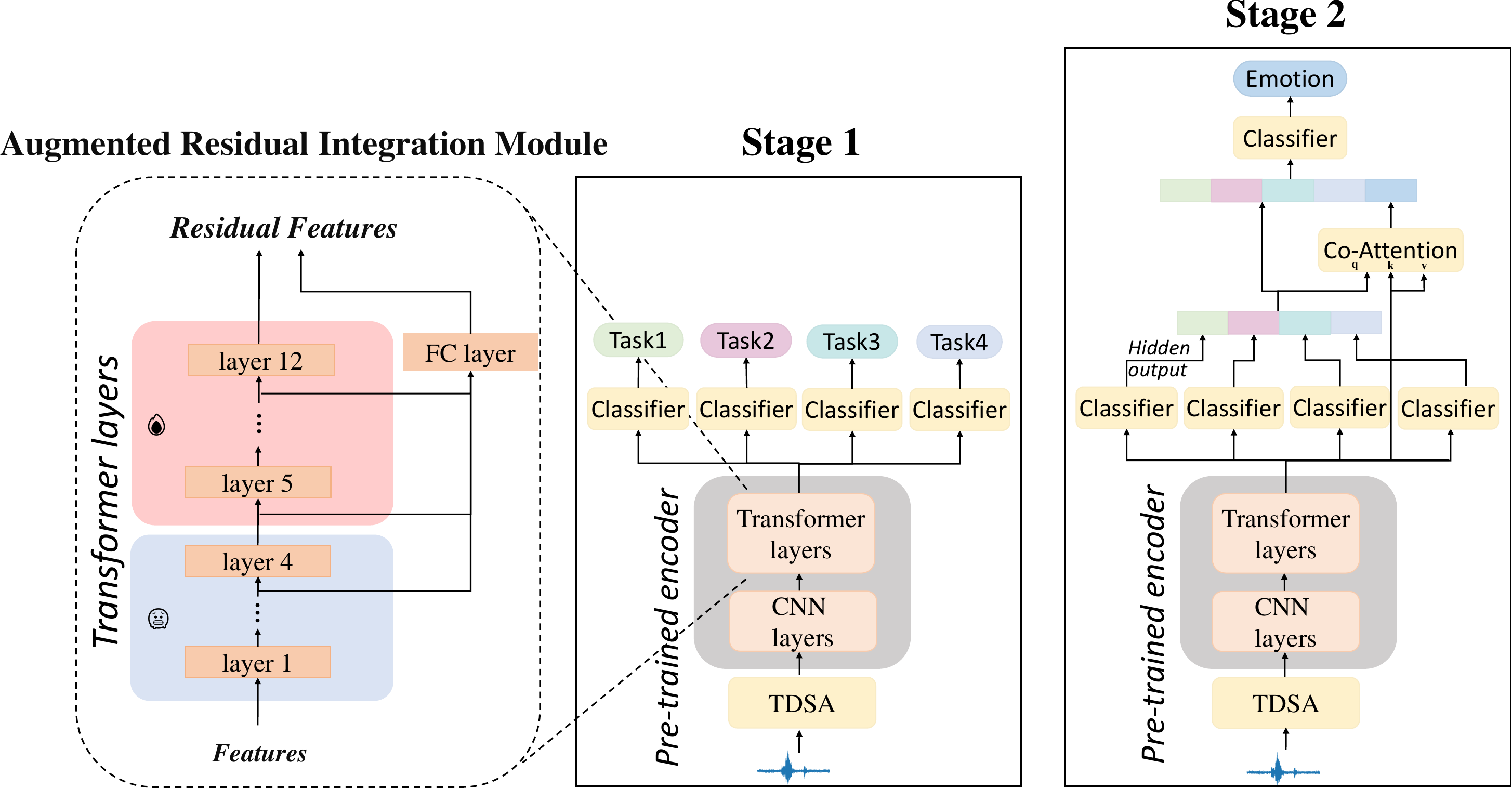}  
  \caption{The proposed model is trained in two stages, the first stage trains the auxiliary tasks and the second stage trains the SER with the auxiliary task information. Note: All Transformer layers are fine-tuned in stage 1, but the first four layers are frozen in stage 2.}  
  \label{fig1}  
\end{figure*}

\section{Proposed Method}



\subsection{Model Architecture}
We propose an end-to-end model that takes speech as raw waveform input and outputs predicted emotions. Figure \ref{fig1} shows the model architecture.

\subsubsection{Time Domain Spectral Augmentation}
Initially, raw waveforms undergo Time Domain Spectral Augmentation (TDSA) for data augment. This module applies speed perturbation by resampling the waveform at varying rates (80, 100, and 120), thereby improving neural model robustness during training. Notably, TDSA is utilized exclusively in the training phase.

\subsubsection{Feature Encoder}
To obtain high-level contextual representations of speech, pretrained SSL-based models were utilized as the original waveform encoder. Three representative transformer-based SSL models—Wav2Vec-2.0-base, HuBert-base \cite{hsu2021hubert}, and WavLM-base \cite{chen2022wavlm}—were selected to validate the generality of the proposed method. These models, all based on self-supervised learning, aim to extract rich representations from raw audio without requiring large amounts of labeled data. Furthermore, all selected models were pretrained on 960 hours of LibriSpeech and share a similar architecture, consisting of: 1) convolutional layers that extract deep features from raw audio, and 2) twelve Transformer layers that capture contextual information from the output of the CNN layers.

Since splitting MTL into stages effectively avoids gradient conflicts and allows the model to better adapt to the current stage of the task \cite{twostageMTL}, a two-stage fine-tuning strategy was adopted for training. In the first stage, the convolutional layers were frozen, and all Transformer layers were fine-tuned to embed auxiliary task information. In the second stage, to maintain stability for tasks that require lower-level features, the last 8 Transformer layers were further fine-tuned while keeping the convolutional layers and the first 4 Transformer layers frozen. By freezing the first a few layers, the emotion recognition task was able to benefit from the information learned in the first stage.


\subsubsection{Multi-task Learning}
As described in the introduction, all available metadata in IEMOCAP were incorporated to facilitate feature extraction. In this work, we incorporated gender, speaker, speech style (impromptu vs. scripted) and Automatic Speech Recognition (ASR) tasks as auxiliary tasks in the MTL model.

For all auxiliary tasks except ASR, as well as SER, cross-entropy loss function was used. 

For ASR, the Connectionist Temporal Classification (CTC) loss function \cite{graves2006connectionist} was used, as it is particularly suited for the sequence-to-sequence learning process inherent in speech recognition.

In the first stage of training, the weighted sum of the loss functions for all auxiliary tasks was minimized:
\begin{equation}
\mathcal{L}_{\mathrm{stage1}} = \alpha\mathcal{L}_{\mathrm{aux\_task1}} + \beta\mathcal{L}_{\mathrm{aux\_task2}} + \gamma\mathcal{L}_{\mathrm{aux\_task3}}+  ...
\end{equation}
where $\alpha,\beta,\gamma$ are hyper-parameters.

In the second stage of training, the following loss function was minimized:
\begin{equation}
\mathcal{L}_{\mathrm{stage2}} = \mathcal{L}_{\mathrm{emotion}}
\end{equation}
where $\mathcal{L}_{\mathrm{emotion}}$ is the loss function of SER.

\subsubsection{Augmented Residual Integration Module (ARI)}
In traditional multi-task learning models, the simple weighted sum method often has shortcomings such as information dilution, where mixing information from all layers can lead to the loss of particular features, and increased complexity, as directly summing outputs from different layers may introduce conflicts, making optimization more challenging.

In order to enhance the feature integration and utilization capabilities of the model, the Augmented Residual Integration (ARI) module was introduced for the Transformer layers in the encoder of transformer-based SSL models. This design is inspired by previous research \cite{chen2022wavlm} on WavLM, which identified that different Transformer layers exert significant effects on different tasks. For instance, layer 1 and layer 4 are especially effective for extracting speaker information, layer 8 through layer 10 are more proficient for emotion recognition, and layer 11 contains the majority of the textual information. The detailed formula is presented as follows:
\begin{equation}
F_{\text{ARI}} = \left[ F_{12} ; W^T \begin{bmatrix} F_1 ; F_2 ; \ldots ; F_{11} \end{bmatrix} \right]
\end{equation}
where $F_i$ denotes the output of the $i$-th Transformer layer;
$W \in \mathbb{R}^{1 \times 11}$ denotes weighted matrix.

In contrast to merely weighting and summing all layers, the ARI module selectively integrates the output of the initial 11 layers and connects it to the output of the last (12-th) layer. This design enables the capture of a more expansive range of task-relevant feature information, while the output of the last layer provides high-level semantic information that is essential for complex tasks.

It is postulated that this module design can be better adapted to different feature-demanding tasks such as gender recognition (GR), speaker recognition (SpkR), ASR, etc. The results of our ablation experiments in \ref{comb} demonstrate that the ARI module outperforms traditional weighted sum method in terms of performance, providing a more robust and task-specific feature representation that is effective across all tasks. Moreover, the results in \ref{result} demonstrate that the proposed method can be generalized to all transformer-based SSL models.

\subsubsection{Co-attention Module}
The Co-attention module integrates the output from the ARI module with the hidden layers of the auxiliary tasks. Recognizing the auxiliary tasks' synergistic impact on Speech Emotion Recognition, we leverage their interplay to guide feature weighting, thereby generating weighted emotion features. Subsequently, these weighted emotion features are combined with the hidden layer outputs from the auxiliary tasks. This composite input is then fed into the SER classifier to generate the predicted labels $\hat{y}$.

\section{Experimental Setup}
\subsection{Data preparation}
In this study, the Interactive Emotional Dyadic Motion Capture (IEMOCAP) dataset was utilized to validate the effectiveness of the proposed method \cite{IEMOCAP2008}. The dataset contains 12 hours of recordings from 10 speakers, with each conversation featuring a male and a female speaker, including both scripted and improvised segments. It covers five emotion labels: neutral, happy, angry, sad, and excited.

Regarding data preparation, to maintain consistency with current state-of-the-art (SOTA) research \cite{shen2023mingling,ye2023temporal,chen2023dst,chen2023dwformer,he2023multiple,twostageMTL}, the excited class was merged with the happy class in the IEMOCAP dataset, resulting in four primary emotion categories: neutral (1,708 samples), happy (1,636 samples), angry (1,103 samples), and sad (1,084 samples). The 5-fold cross-validation with speaker-independent constraint (with one pair of speakers per fold) was applied. The results were compared using weighted accuracy (WA) and unweighted accuracy (UA).

It is crucial to highlight that all metadata within the dataset were utilized, as they are integral to this research. The metadata in IEMOCAP includes speaker gender, speaker ID, speech style, and audio transcription.

\subsection{Implementations}
The proposed model was implemented using PyTorch and SpeechBrain \cite{paszke2019pytorch, ravanelli2021speechbrain}. Building on previous research \cite{twostageMTL,ZhaoJ2022_JSTSP}, three representative transformer-based SSL models—Wav2Vec-2.0-base, HuBert-base, and WavLM-base—were selected. All three models are pre-trained on 960 hours of LibriSpeech \cite{panayotov2015librispeech}. These models shared a similar architecture including a CNN feature extractor for mapping raw audio to latent representations, as well as twelve Transformers layers for learning contextual information. All sentences in each batch were padded to ensure compatibility with the input length requirements of the SSL models. During training, the CNN feature extractor was frozen, and the Transformer layers were fine-tuned for 100 epochs. To reduce the dimensionality of features, mean pooling and two FC layers with dropout were employed for SER. The downstream models used a learning rate of 10e-4, while the SSL models used a learning rate of 10e-5. The batch size was set to 4. For the hyperparameter settings of the metadata auxiliary tasks, IEMOCAP includes gender, speaker, speech style (impromptu vs. scripted), and Automatic Speech Recognition (ASR) tasks. Each task was assigned a hyperparameter independently. The hyperparameter was subject to change with different encoders.

\section{Experimental Results and Analysis}

\subsection{Evaluation Results}\label{result}

Table \ref{t1} presents a comparison between the proposed method, baseline models, and methods with leading performance. First, compared to the baseline models which leverage the base version of each SSL encoder followed by a fully connected layer as classifier, the proposed method, using the corresponding encoder of the baseline models, achieves a significant improvement, with an average increase of 4.09\% in UA and 3.90\% in WA. Second, under speaker-independent conditions and using different SSL encoders, the proposed method consistently outperforms the leading studies, achieving state-of-the-art results with improvements ranging from 0.54\% to 1.64\% in UA and from 0.9\% to 1.39\% in WA. These results demonstrate the effectiveness and superiority of the proposed method in enhancing the performance of pretrained transformer-based SSL models.

\begin{table}[htb]
  \caption{Comparison with known leading performance levels systems}
  \label{t1}
  \centering
    \scalebox{0.8}{
  \begin{tabular}{l|c|c|c}
    \toprule
    \toprule
    \textbf{IEMOCAP} &\textbf{CV} & \textbf{UA}&\textbf{WA} \\
    \midrule
    \textbf{Prior-art}\\

    Shen et al. \cite{shen2023mingling} &5-fold& $72.70$ & $72.10$  \\
    Ye et al. \cite{ye2023temporal} &10-fold& $72.50$ & $71.65$ \\
    Chen et al. \cite{chen2023dwformer} &5-fold &$73.90$ &$72.30$ \\
    He et al. \cite{he2023multiple}&5-fold&$74.25$&$73.80$\\
    Gao et al. \cite{twostageMTL} &5-fold& $76.10$ & $74.94$  \\
    \midrule
    \textbf{Our baseline}\\
    Wav2Vec-2.0&5-fold&$72.38$&$71.14$ \\
    HuBert  &5-fold& $72.91$ & $72.08$ \\
    WavLM  &5-fold& $73.79$ & $73.12$ \\
    \midrule
    \textbf{Proposed}\\
   Ours(Wav2Vec-2.0)&5-fold& $76.64$ &  $75.84$ \\
   Ours(HuBert) &5-fold& $76.97$ &  $75.88$ \\
   \textbf{Ours(WavLM)} &\textbf{5-fold}& $\textbf{77.74}$ &  $\textbf{76.33}$ \\
    \toprule
    \toprule
  \end{tabular}}
\end{table}

\subsection{Ablation Study on Highlighted Modules}

Table \ref{t2} presents the results of the ablation study on the ARI and Co-attention modules. Performance improvements are observed across the three transformer-based encoders with both metrics when the ARI and Co-attention modules are included. Specifically, UA increases by 0.44\% and WA by 0.61\% with the addition of the ARI module. When the Co-attention module is added, there is an increase of 0.64\% in UA and 0.82\% in WA. This indicates that the ARI module's capability to preserve multi-level features and the Co-attention module's ability to capture contextual information from tasks both contribute to performance enhancement.

\begin{table}[htb]
  \caption{ABLATION STUDY of ARI and Co-attention module}
  \label{t2}
  \centering
    \scalebox{0.8}{
  \begin{tabular}{l|c|c|c|c|c|c}
    \toprule
    \toprule
    \multirow{2}{*}{\diagbox[dir=SE]{\textbf{Models}}{\textbf{Metric}}{\textbf{Encoder}}} &\multicolumn{2}{c|}{\textbf{Wav2Vec-2.0}} & \multicolumn{2}{c|}{\textbf{HuBert}}&\multicolumn{2}{c}{\textbf{WavLM}} \\
    \cmidrule{2-7}
    &\textbf{UA}&\textbf{WA} &\textbf{UA}&\textbf{WA} &\textbf{UA}&\textbf{WA} \\
    \midrule
    \textbf{Baseline} &$72.38$&$71.14$ & $72.91$ & $72.08$ & $73.79$ & $73.12$ \\
    \midrule
    \multicolumn{1}{l}{\textbf{Proposed}}\\
    Ours w/o ARI\&Co  & $75.09$ & $73.69$ & $75.92$ &$74.90$ &$76.42$&$75.59$ \\
    Ours w/o ARI  & $75.94$ & $74.99$  & $76.06$ &$75.08$ &$76.75$&$75.94$\\
    Ours w/o Co  & $76.30$ & $75.49$ & $76.35$ &$75.43$ &$76.71$&$75.71$\\
   \textbf{Our}& $\textbf{76.64}$ &  $\textbf{75.84}$ &$\textbf{76.97}$&  $\textbf{75.88}$ &  $\textbf{77.74}$ &$\textbf{76.33}$  \\
    \toprule
    \toprule
  \end{tabular}}
\end{table}

However, when both the ARI and Co-attention modules are included together, the increase in UA reaches 1.31\%, and WA improves by 1.29\%, both surpassing the performance gains of adding ARI or Co-attention individually. This demonstrates that the ARI module effectively preserves multidimensional, high-quality auxiliary information, while the Co-attention module excels at utilizing multidimensional and contextual information from metadata-related auxiliary tasks, working in a complementary manner.

\subsection{Analysis of ARI modules} \label{ab}


Table \ref{t3} presents the performance of the weighting and summing all 12 layers (weighted sum) method and the ARI module on metadata-related tasks, alongside the final results after applying the two-stage fine-tuning framework. GR, StyR and SpkR correspond to the tasks of gender recognition, speech style recognition and speaker recognition, respectively. The results in Table  \ref{t3} show that, compared to the weighted sum method, the ARI module demonstrates superior performance across every task.

\begin{table}[htb]
  \caption{Auxiliary Tasks and Final Results}
  \label{t3}
  \centering
    \scalebox{0.80}{
  \begin{tabular}{c|c|c|c|c|c|c}
    \toprule
    \toprule
    \diagbox{\textbf{Encoder}}{\textbf{Task}}  & \textbf{CER} & \textbf{WER} & \textbf{GR acc} & \textbf{StyR acc} & \textbf{UA} & \textbf{WA}\\
    \midrule
    \multicolumn{1}{l}{\textbf{Wav2Vec-2.0}}\\
    Weighted sum & 14.49 & 35.78 & 97.91 & 84.75 & 75.59 & 74.66   \\
    ARI & \textbf{13.81} & \textbf{33.64} & \textbf{98.46} & \textbf{85.20} & \textbf{76.64} & \textbf{75.84}   \\
    \midrule
    \multicolumn{1}{l}{\textbf{HuBert}}\\
    Weighted sum &21.03 &	55.69&	97.57	&84.95		&74.97&	74.08  \\
    ARI & \textbf{13.64} & \textbf{32.19} & \textbf{98.32} & \textbf{86.95} & \textbf{76.97} & \textbf{75.88}   \\
    \midrule
    \multicolumn{1}{l}{\textbf{WavLM}}\\
    Weighted sum &17.35&	47.47	&98.07&	85.83&		75.53	&74.73  \\
    ARI & \textbf{11.56} & \textbf{27.98} & \textbf{98.71} & \textbf{87.34} & \textbf{77.74} & \textbf{76.33}   \\
    \toprule
    \toprule
    \multicolumn{7}{l}{SpkR accuracy for IEMOCAP is not shown as train and test set speakers are }\\
    \multicolumn{7}{l}{mutually exclusive, leading to 0 accuracy. }\\
  \end{tabular}}
\end{table}

Specifically, for ASR (automatic speech recognition) tasks, the improvement is most significant, with a relative average improvement of 24.40\% in CER and an average improvement of 29.75\% in WER. However, for gender recognition and style recognition tasks, the relative average improvement is only 0.66\% and 1.55\%, respectively. The substantial improvement in ASR tasks supports the observation that the higher the weight assigned to the ASR task during fine-tuning, the more likely the model is to achieve better performance.

On the IEMOCAP dataset, the four shown metrics exhibit an average improvement of 5.41\% over the weighted sum method. Furthermore, the final results after applying the two-stage fine-tuning method also maintain the same trend.

After validating in three representative transformer-based encoders with 4 different auxiliary tasks, it is safe to conclude that the ARI module is more effective and robust than the weighted sum approach in selecting features better suited for different feature-demanding tasks, thereby providing the Co-attention module with higher-quality auxiliary information derived from multimodal metadata.

\subsection{Analysis of Auxiliary Task Combinations} \label{comb}

In our investigation, we integrated various auxiliary metadata into the proposed SER model. Ablation studies in Table \ref{t4} examine performance under different auxiliary task configurations. A clear improvement in model efficacy correlates with the incremental integration of auxiliary tasks. Notably, in IEMOCAP, speaker identification accuracy was zero due to the speaker-independent setup (mutually exclusive speakers in training and test sets). However, integrating speaker identification still improved performance in the multi-task learning (MTL) framework (e.g., from speaker + gender to speaker + gender + style), showing that the Co-attention module effectively captures relationships between auxiliary tasks, thereby contributing to the primary task.

\begin{table}[htb]
  \caption{Ablation study of auxiliary tasks on IEMOCAP}
  \label{t4}
  \centering
\scalebox{0.80}{
  \begin{tabular}{c|c|c||c|c||c|c}
    \toprule
    \toprule
    \multicolumn{1}{c|}{\textbf{Encoder}} & \multicolumn{2}{c||}{\textbf{Wav2Vec-2.0}}  & \multicolumn{2}{c||}{\textbf{HuBert}}  & \multicolumn{2}{c}{\textbf{WavLM}} \\
    \midrule
    \diagbox{\textbf{Task}}{Metric} & \textbf{UA}&\textbf{WA} & \textbf{UA}&\textbf{WA} & \textbf{UA}&\textbf{WA}\\
    \midrule
    SER only & $72.57$ & $71.91$ &$73.88$   &$73.35$   &$74.65$  &$73.42$  \\
    \midrule
    SER+gender & $67.91$ & $66.64$&$72.00$   &$71.37$   &$74.11$  &$73.52$   \\
    SER+speaker & $71.55$ & $70.89$  &$70.81$   &$70.24$   &$74.67$  &$73.65$ \\
    SER+style & $73.68$ & $72.72$  &$74.17$   &$73.32$   &$75.62$  &$74.22$ \\  
    SER+ASR & $75.73$ & $74.94$  &$75.91$   &$74.68$   &$76.60$  &$75.92$ \\
    \midrule
    SER+gender+speaker & $72.35$  & $71.35$ &$71.35$   &$70.73$   &$73.99$  &$72.66$ \\
    SER+gender+style & $74.39$ & $73.71$ &$74.56$   &$73.55$   &$74.90$  &$73.79$  \\
    SER+gender+ASR & $76.40$ & $75.40$  &$76.40$   &$75.23$   &$77.06$  &$76.48$ \\
    SER+speaker+style & $73.89$& $73.09$   &$73.28$   &$72.17$   &$74.82$  &$73.43$ \\
    SER+speaker+ASR & $75.87$& $75.10$  &$76.48$   &$75.23$   &$77.14$  &$76.39$ \\
    SER+style+ASR & $76.62$& $75.78$ &$76.09$   &$75.51$   &$77.02$  &$76.27$ \\
    \midrule
    SER+gender+speaker+style & $72.83$ & $72.09$ &$73.54$   &$72.26$   &$75.06$  &$73.48$ \\
    SER+gender+speaker+ASR & $76.49$ & $75.79$  &$76.79$   &$75.82$   &$77.35$  &$76.20$ \\
    SER+gender+style+ASR & $76.61$ & $75.69$ &$76.30$   &$75.12$   &$77.24$  &$76.29$ \\
    SER+speaker+style+ASR & $76.39$ & $75.55$&$76.86$   &$75.62$   &$77.57$  &$76.13$  \\
    \midrule
    SER+gender+speaker+style+ASR & $\textbf{76.64}$ & $\textbf{75.84}$ &\textbf{76.97} &\textbf{75.88}   &\textbf{77.74}  &\textbf{76.33} \\
    \bottomrule
    \bottomrule
  \end{tabular}}
\end{table}

This also explains the performance drop when fewer auxiliary tasks are used—limited modalities restrict contextual information, and some metadata (e.g., gender) may be weakly related to SER. This can be mitigated by freezing the ARI module in Stage 1. Encouragingly, with enough modalities, the multi-task learning (MTL) framework consistently outperforms SER-only model. Additionally, in the MTL framework, each additional modality, regardless of its relevance to the primary task, improves performance, further demonstrating the Co-attention module's effectiveness on capturing and utilizing the relationships between all auxiliary tasks and their associated metadata.

\section{Conclusion and Limitations}

In this paper, we introduce a metadata-enhanced approach for transformer-based SSL models through the ARI and Co-attention modules, based on a two-stage fine-tuning framework. The experiment highlights the ARI module's capability to provide more robust and task-specific feature selection, enabling the extraction of higher-quality and more diverse auxiliary information from metadata. The ablation study reveals a promising correlation between auxiliary tasks and model performance, demonstrating the Co-attention module's ability to efficiently capture and utilize the relationships between various metadata auxiliary tasks. The ARI and Co-attention modules exhibit complementary strengths in preserving and utilizing multimodal information, unveiling a methodology that enhances the model's ability to process multidimensional information based on transformer-based SSL models. Moreover, the fact that this method outperforms SOTA across different SSL encoders under pre-trained base models and a speaker-independent setup further validates its effectiveness and superiority. In future work, the potential improvements with larger pretrained models and the generalizability of this methodology to other tasks in the speech domain will be further explored.

\bibliographystyle{IEEEtran}
\bibliography{mybib}

\end{document}